\documentclass[pre,aps]{revtex4-1}
\usepackage{epsfig,amsmath,graphicx,amssymb,overpic}
\def\be{\begin{equation}}
\def\ee{\end{equation}}
\def\bee{\begin{eqnarray}}
\def\ene{\end{eqnarray}}
\def\bes{\begin{subequations}}
\def\ees{\end{subequations}}

\begin{document}
 \title{\large Localized analytical solutions and parameters analysis in the nonlinear dispersive Gross-Pitaevskii
 mean-field GP$(m, n)$ model with space-modulated nonlinearity and potential}
\author{\it  Zhenya Yan\vspace{0.1in}~\footnote{Address for correspondence: Zhenya Yan, Key Laboratory of Mathematics Mechanization, Institute of Systems Science, AMSS, Chinese Academy of Sciences,
Beijing 100190, People's Republic of China; e-mail: zyyan@mmrc.iss.ac.cn}}
\baselineskip=20pt

\date{\vspace{0.1in} June 2013, Stud. Appl. Math.  {\bf 132}, 266 (2014)}

\begin{abstract}

\noindent\rule{140mm}{0.5pt}

 \vspace{0.1in} \baselineskip=20pt

\noindent The novel nonlinear dispersive Gross-Pitaevskii (GP)
mean-field model with the  space-modulated nonlinearity and potential (called GP$(m, n)$
equation) is investigated in this paper. By using self-similar
transformations and some powerful methods, we obtain some families of novel envelope
compacton-like solutions spikon-like solutions to the GP$(n, n) \
(n>1)$ equation. These solutions possess abundant localized
structures because of infinite choices of the self-similar
function $X(x)$. In particular, we choose $X(x)$ as the Jacobi
amplitude function ${\rm am}(x,k)$ and the combination of linear
and trigonometric functions of space $x$ so that the novel localized
structures of the GP$(2,2)$ equation are illustrated, which are
much different from the usual compacton and spikon solutions
reported.  Moreover, it is shown that GP$(m,1)$ equation with
linear dispersion also admits the compacton-like solutions for the case
$0<m<1$ and spikon-like solutions for the case $m<0$.

\noindent\rule{140mm}{0.5pt}
\end{abstract}

\maketitle


\baselineskip=20pt

\centerline{\bf\large 1. \ Introduction}\vspace{0.10in}

Soliton plays a more and more important role and has many
applications in the field of nonlinear science such as plasma physics,
nonlinear optics, Bose-Einstein condensation, and finance, etc.
The generation of soliton is due to the balance between (linear) dispersion and the nonlinear interaction~\cite{soli}.
Since soliton was coined by Zabusky and Kruskal in 1965~\cite{Soliton}, many new types of solitons have been reported
such as optical solitons, breather solitons, dromion solutions,
peakons, compactons, rogons, etc~\cite{Op, Op1, Op2, Op3,S2, yan10, R1}. It is still
an interesting subject to investigate various of exact
analytical solutions, in particular solitons, of nonlinear physical
equations.

The compacton was first presented in the study of the KdV equation
with nonlinear dispersion (called K$(m,n)$ equation) twenty years ago~\cite{R1, R2, R22},
and it was shown that the K$(n,n)$
equation admitted a new kind of solitons, called compactons, which
usually are described by powers of trigonometric functions in its
one minimum period and exist in nonlinear wave equations with
nonlinear dispersion. While compactons are the essence of the
focusing branch, spikes, peaks and cusps are the hallmark of the
defocusing branch which also supports the motion of kinks. The
defocusing branch was found to give rise to solitary patterns
having infinite slopes or cusps~\cite{R1}-\cite{Yan62}. Up to now
many different types of compactons were also presented, containing
the discrete compactons~\cite{D1, D2, D22, D3, D33}, breather
compactons~\cite{BC, BC2, BC3}, elliptic compactons~\cite{Yane}, envelope
compactons~\cite{Yan6, Yan62}, etc. Moreover, it was found that nonlinear
dispersion is not necessary condition to possess compactons and
solitary patterns for nonlinear wave equation~\cite{Yan, Yan1, Yan6, Yan62}. Recently,
we have presented the new nonlinear dispersive K$(m, n)$ model with
variable coefficients and investigated its some solutinos~\cite{Yank}.

The one-dimensional Gross-Pitaevskii (GP) mean-field equation
\bee
 i\hbar\frac{\partial}{\partial t}\psi=\left[-\frac{\hbar^2}{2 M}\frac{\partial^2}{\partial x^2}
  +V_{\rm ext}(x)+g_{\rm 1D}|\psi|^2\right]\psi, \label{gp}
  \ene
is a very important model to describe the static and dynamical properties of a
Bose-Einstein condensation (BEC)~\cite{BEC,BEC2,BEC3,BEC4}, where $\psi\equiv\psi(x,t)$ is the condensate wavefunction, $M$ is the atmoic mass,
$V_{\rm ext}(x)$ denotes the external potential and is usually chosen as the harmonic potential and optical lattice potential, $g_{\rm 1D}=2\hbar^2\omega_{\bot}a_s/M$ stands for
the effective one-dimensional coupling strength with $\omega_{\bot}$ being the transverse confining frequency,
and $a_s$ being the $s$-wave scattering length ($a_s>0\, (<0)$ corresponding to the repulsive (attractive) interaction), in which $a_s$ is a function of the magnetic field $B$ in the form~\cite{fr1,fr2}
$$a_s(B)=a_0\left(1-\frac{\Delta}{B-B_0}\right),$$
where $a_0$ denotes the background scattering length, which is the scattering length associated with the background potential, the parameter $B_0$ stands for the resonance position, and the parameter $\Delta$ is the
resonance width.

Up to now, various types of the generalized GP equation with space- and time-modulated coefficients have been
reported~\cite{GP,GP1, GP2, GP3, GP4} such as space- and time-modulated potential and nonlinearity, and the higher-degree nonlinearities. Recently, we introduced and studied the novel nonlinear Schr\"odinger
equation with nonlinear dispersion $\partial_x^2(|\psi|^{n-1}\psi)$ and constant coefficients
(called NLS$(m,n)$ equation)~\cite{Yan6}
\begin{eqnarray}
\label{NLS}
 i\frac{\partial \psi}{\partial t}
  =-\frac{\partial^2}{\partial x^2}(|\psi|^{n-1}\psi)+\gamma|\psi|^{m-1}\psi, \ \
 (\gamma=\pm 1)
 \end{eqnarray}
and two generalized higher-order nonlinear Schr\"odinger equations with nonlinear dispersion  and constant coefficients (called GNLS$(m,n)$ equations)~\cite{Yan62}
\begin{eqnarray}
\label{NLS}
 i\frac{\partial \psi}{\partial t}+a\frac{\partial^2}{\partial x^2}(|\psi|^{n-1}\psi)+b|\psi|^{m-1}\psi+ic\frac{\partial^3}{\partial x^3}(|\psi|^{p-1}\psi)+id\frac{\partial}{\partial x}(|\psi|^{q-1}\psi)=0,
 \end{eqnarray}
\begin{eqnarray}
\label{NLS}
 i\frac{\partial \psi}{\partial t}+a\frac{\partial^2}{\partial x^2}(|\psi|^{n-1}\psi)+b|\psi|^{m-1}\psi+ic\frac{\partial^3}{\partial x^3}(|\psi|^{p-1}\psi)+ih|\psi|^{q-1}\frac{\partial}{\partial x}\psi=0,
 \end{eqnarray}
and obtained their some types of envelope compactons and spikons for some different parameters $m,\,n,\, p,\,q$ ~\cite{Yan6, Yan62}, where $m,\,n,\, p,\,q,\, a,\,b,\,c,\, d,\,$ and $h$ are real-valued constants. To our knowledge, compacton-like
and spikon-like solutions of nonlinear complex dispersive wave equations with varying coefficients were not reported before.

In this paper, we extended the ideas~\cite{Yank, Yan6, Yan62} to the GP  equation and introduce the nonlinear dispersive GP equation with
space-modulated potential and nonlinearity (called GP$(m,n)$ equation) such that novel localized
solutions are found. These solution profiles are very
different from the usual compacton and spikon solutions.  The rest
of this paper is organized as follows. In Section 2, we
 introduce the nonlinear dispersive GP model with varying
 coefficients, which is called the GP$(n,n)$ equation. In Section 3, we obtained
 self-similar solutions including compacton-like and spikon-like solutions of
 GP$(n,n)$ and GP$(m,1)$ equations. We analyze the localized
 solutions for the chosen function $X(x)$ to be the Jacobi amplitude
 function ${\rm am}(x,k)$ and the combination of linear
and trigonometric functions of $x$ in Section 4. Finally,
 some conclusions are given in last section.\\

\centerline{\bf\large 2. \ Nonlinear dispersive GP model with
space-modulated potential}
\centerline{\bf\large and nonlinearity: GP$(m,n)$ equation} \vspace{0.10in}

To understand the role of nonlinear dispersion in the one-dimensional dimensionless  GP
mean-field model arising from Bose-Einstein condensates, we introduce and study the dimensionless nonlinear dispersion GP equation by replacing the linear  dispersion $\partial_x^2 \psi$ with nonlinear dispersion $\partial_x^2(|\psi|^{n-1}\psi)$ and changing the nonlinear term, described
by
 \begin{eqnarray}
 i\frac{\partial \psi}{\partial t}
  =-\frac{\partial^2}{\partial x^2} (|\psi|^{n-1}\psi)+v(x)\psi+g(x)|\psi|^{m-1}\psi, \qquad (n\geq 1)
  \label{GP}
 \end{eqnarray}
which is simply referred to as the GP$(m,n)$ equation, where
$\psi=\psi(x,t)$ is a complex field, $m$ and $n$ are real-valued
parameters, $v(x)$ is  a linear (trap) potential, and $g(x)$
describes the spatial modulation of the nonlinearity. GP$(m,n)$ equation (\ref{GP}) contains many types of nonlinear wave equations. If $v(x)\equiv 0$ and $g(x)$ is a constant, then Eq.~(\ref{GP}) becomes the NLS$(m,n)$ equation (\ref{NLS}). Though Eq.~(\ref{GP}) is in fact the generalized GP equation with the linear dispersion for the case $n=1$, but we will investigate its new wave structures such as compacton and spikon solutions not solitary wave structures. In particular, i) for the case $n=1$ and $m=3$,  the GP$(3,1)$ equation reduces to the usual GP
equation with space-modulated coefficients
 \begin{eqnarray}
 i\frac{\partial \psi}{\partial t}
  =-\frac{\partial^2 \psi}{\partial x^2}+v(x)\psi+g(x)|\psi|^2\psi,
  \label{GP3}
 \end{eqnarray}
 whose periodic wave solutions and solitary wave solutions were
obtained for different potentials and nonlinrarities~\cite{GP,GP1,GP2,GP3,GP4}; ii)  for the case $n=1$ and $m>1$, the GP$(m,1)$ equation becomes the
generalized GP model~\cite{GPg}
 \begin{eqnarray}
 i\frac{\partial \psi}{\partial t}
  =-\frac{\partial^2 \psi}{\partial x^2}+v(x)\psi+g(x)|\psi|^{m-1}\psi,
  \label{GPg}
 \end{eqnarray}
iii)  for the case $n=m=1$, the GP$(1,1)$ equation becomes the linear GP (NLS) model with the potential $v(x)+g(x)$:
 \begin{eqnarray}
 i\frac{\partial \psi}{\partial t}=-\frac{\partial^2 \psi}{\partial x^2}+[v(x)+g(x)]\psi.
  \label{GPl}
 \end{eqnarray}

 In what follows, we will focus on compacton-like solutions and
spikon-like solutions of GP$(m,n)$ equation except for the above-mentioned three cases by using self-similar
transformations and some ansatze. \\

\centerline{\bf\large 3. \ General theory and self-similar
solutions} \vspace{0.10in}

In general, Eq.~(\ref{GP}) is not integrable for the case $n\not=1,\, m\not=2$. It is difficult to solve directly Eq.~(\ref{GP}). We need reduce Eq.~(\ref{GP}) to some equations solving easily. Eq.~(\ref{GP})
may possess many types of similarity reductions by using the symmetry analysis (see, e.g.,~\cite{lie, lie2}). Our goal is to reduce the solutions of GP$(m,n)$ equation
(\ref{GP}) to those of the stationary Gross-Pitaevskii equation
with nonlinear dispersion (called SGP$(m,n)$ equation) \bee
 \mu\Phi(X)=-\frac{d^2\Phi^n(X)}{dX^2}+G\Phi^m(X),
 \label{OGP}
 \ene
where $\Phi(X)$ is the real-valued stationary field, $X\equiv X(x)$ is an
unknown function of space $x$ to be determined, $\mu$ is the real eigenvalue of
the nonlinear wave equation (\ref{OGP}), and $G$ is a real
coefficient of the nonlinearity. Eq.~(\ref{OGP}) is a complicated nonlinear ordinary differential equation.
When $n=1,\, m=3$, Eq.~(\ref{OGP}) is the stationary nonlinear Schr\"odinger equation or stationary Gross-Pitaevskii equation without a potential
$\mu\Phi(X)=-\frac{d^2\Phi(X)}{dX^2}+G\Phi^3(X)$, which admits the bright ($G<0$) and dark ($G>0$) solitary wave solutions~\cite{GP3}.
We explore the following self-similar transformation (the stationary solutions)
  \bee \psi(x,t)=\rho(x)e^{i\varphi(t)}\Phi[X(x)], \label{Tr} \ene
to the GP$(m,n)$ equation (\ref{GP}) such that we have the nonlinear differential equation:
\bee\begin{array}{l}
\varphi_t(t)\rho(x)\Phi(X)-n\rho^n(x)X_x^2\Phi^{n-1}(X)\Phi_{XX}(X)-n(n-1)\rho^n(x)X_x^2\Phi^{n-2}(X)\Phi_{X}^2(X)
 \vspace{0.15in}\cr
 \quad -n\rho^{n-1}(x)\rho_{xx}(x)\Phi^n(X)-2n^2\rho^{n-1}(x)\rho_x(x)X_x\Phi^{n-1}(X)\Phi_{X}(X)-n\rho^n(x)X_{xx}\Phi^{n-1}(X)\Phi_{X}(X)
  \vspace{0.15in}\cr
  \quad -n(n-1)\rho^{n-2}(x)\rho_x^2(x)\Phi^n(X)+v(x)\rho(x)\Phi(X)+g(x)\rho^m(x)\Phi^m(X)=0,
 \end{array} \label{sys}
 \ene
where the subscripts denote the partial derivative with respect to the related variables,
 $\rho(x)$ denotes the amplitude of the wave function, and $\varphi(t)$ is the phase,

It follows from Eq.~(\ref{sys}) that all other terms are functions of space $x$ except for the term $\varphi_t(t)\rho(x)\Phi(X)$. Thus we require that the function $\varphi_t(t)$ should be a non-zero constant (e.g, $\varphi_t=\omega={\rm const}$). Since we require that $\Phi(X)$ satisfies Eq.~(\ref{OGP}), thus we
balance the coefficients of the related terms to obtain the following two possible systems
in these unknown functions $\rho(x),\ g(x),\ v(x)$, and $\varphi(t)$ for different types of parameters $m$ and $n$:

 \vspace{0.1in}
\noindent {\it System I} : \ for the case $n=m$.
   \bee \begin{array}{l}
  \varphi_t(t)=\omega, \vspace{0.1in} \cr
  [\rho^{2n}(x)X_x]_x=0, \vspace{0.1in} \cr
   v(x)=-\mu\rho^{n-1}(x)X_x^2-\omega, \vspace{0.1in} \cr
  g(x)=\displaystyle\frac{[\rho^n(x)]_{xx}}{\rho^n(x)}+GX_x^2,
 \end{array}
  \label{eq:5}
  \ene
\noindent {\it System II} : \ for the case $n=1$.
\bee \begin{array}{l}
 \varphi_t(t)=\omega, \vspace{0.1in} \cr
 [\rho^{2}(x)X_x]_x=0, \vspace{0.1in} \cr
  v(x)=\displaystyle\frac{\rho_{xx}(x)}{\rho(x)}-\mu X_x^2-\omega,\vspace{0.1in} \cr
  g(x)=G\rho^{1-m}(x)X_x^2,
  \end{array} \qquad \ \
  \label{eq:6}
  \ene
where $\omega$ is the chemical potential.

The compacton solutions and spikon solutions of the SGP$(m,n)$ equation
(\ref{OGP}) are listed in Table I for differential parameters $G,\
\mu,\ m$ and $n$ by using the direct cosine and sinh-cosh
transformations~\cite{R1, R2, Yan, Yan1, Yan6, Yan62}, in which the first three
solutions are compacton solutions and other ones are spikon
solutions of SGP$(m,n)$ equation. For the first three compacton
solutions in Table I, we require that $|\nu X|\leq \pi/2$, and
$\Phi(X)$ vanishes elsewhere. We can determine, after some
straightforward algebra, the corresponding functions, necessary
for GP$(m,n)$ equation (\ref{GP}) to admit analytical envelope
compacton-like solutions and spikon-like solutions in terms of the
self-similar transformation (\ref{Tr}).

 If we choose $X=X(x)$ as a free function, then one can find that the nonlinearity $g(x)$ and
external potential $v(x)$ depend only on $X(x)$ and two sets of exact solutions of system (I) and (II) are listed
below: \vspace{0.1in}

\noindent {\it Solution I} : \ for the case $n=m.$ \bee\begin{array}{l}
  \varphi(t)=\omega t, \vspace{0.1in} \cr
  \rho(x)=\displaystyle\left(CX_x\right)^{-\frac{1}{2n}},  \vspace{0.1in} \cr
  g(x)=\displaystyle \sqrt{CX_x}\left(\frac{1}{\sqrt{CX_x}}\right)_{xx}+GX_x^2, \vspace{0.1in} \cr
  v(x)=\displaystyle-\mu
  X_x^2\left(CX_x\right)^{\frac{1-n}{2n}}-\omega,\end{array}
  \label{eq:7}
  \ene

\noindent  {\it Solution II} : \ for the case $n=1.$ \bee \begin{array}{l}
  \varphi(t)=\omega t, \vspace{0.1in} \cr \rho(x)=\displaystyle\left(CX_x\right)^{-\frac{1}{2}}, \vspace{0.1in} \cr
  g(x)=\displaystyle GX_x^2\left(CX_x\right)^{\frac{m-1}{2}}, \vspace{0.1in} \cr
  v(x)=\displaystyle\sqrt{CX_x}\left(\frac{1}{\sqrt{CX_x}}\right)_{xx}
   -\mu X_x^2-\omega,\end{array}
  \label{eq:8}
  \ene
where $C\not=0$ is an integration constant, and $CX_x>0$.

 Therefore, in
terms of transformation (\ref{Tr}), we obtain the following two families of novel
analytical solutions with an arbitrary function $X(x)$ of the
GP$(m,n)$ equation (\ref{GP})
 \bee \psi_1(x,t)=\frac{e^{i\omega
 t}}{\sqrt[2n]{CX_x}}\Phi[X(x)],\ \ {\rm for} \ \ n=m>1,
 \label{eq:9} \ene
\bee \psi_2(x,t)=\frac{e^{i\omega
 t}}{\sqrt{CX_x}}\Phi[X(x)],\ \ {\rm for} \ \ n=1,\ m<1,
 \label{eq:10} \ene
where the solutions $\Phi(X(x))$ of the SGP$(m,n)$ equation
(\ref{OGP}) with the different parameters are given in Table I, in which we require that $\mu>0$ for cases $2$ and $3$ and $\mu<0$ for cases $4,\, 5,\, 8,$ and $9$.

\begin{table}
{\small TABLE I. \ Solutions of the SGP$(m,n)$ equation (\ref{OGP})}
\\[1.0ex]
\begin{tabular}{p{1cm}p{1cm} p{2.8cm} p{2.8cm} p{2.8cm} p{2.6cm}} \hline\hline
\\[-2.0ex]
 Case & \centerline{$G$} &
\qquad $A$  & \qquad $\nu$ & \quad $(n,\, m)$ & \qquad $\Phi(X)$  \\ [-2.0ex]  \hline \\[-2.0ex]
 1  & $-1$ & $\ \ \displaystyle-\frac{2n\mu}{n+1}$ & $\ \ \displaystyle \frac{n-1}{2n}$ & $n=m>1$ & $[A\cos^2(\nu X)]^{\frac{1}{n-1}}$
  \vspace{0.1in}\\ [1ex]
 2 & $-1$ & $ \displaystyle-\frac{\mu (m+1)}{2}$ & $\displaystyle\frac{\sqrt{\mu}\,(1-m)}{2}$ & $n=1$ &
   $[A\cos^2(\nu X)]^{\frac{1}{1-m}}$ \vspace{0.1in} \\ [1ex]
 3 & $\ \ 1$ & $\ \ \displaystyle\frac{\mu (m+1)}{2}$ & $\displaystyle\frac{\sqrt{\mu}\,(1-m)}{2}$ & $n=1$ &
 $[A\cos^2(\nu X)]^{\frac{1}{1-m}}$
 \vspace{0.1in}\\ [1ex]
 4 & $-1$ & $\ \ \displaystyle\frac{\mu (m+1)}{2}$ & $\displaystyle\frac{\sqrt{-\mu}\,(1-m)}{2}$ & $n=1,\, m<1$ & $[A\sinh^2(\nu X)]^{\frac{1}{1-m}}$
 \vspace{0.1in}\\
 5 & $-1$ & $\displaystyle-\frac{\mu (m+1)}{2}$ & $\displaystyle\frac{\sqrt{-\mu}\,(1-m)}{2}$ & $n=1,\, m<1$ & $[A\cosh^2(\nu X)]^{\frac{1}{1-m}}$
 \vspace{0.01in} \\ [1ex]
 6  & $\ \ 1$ & $\ \ \displaystyle-\frac{2n\mu}{n+1}$  &  $\ \ \displaystyle\frac{n-1}{2n}$ & $n=m>1$ &  $[A\sinh^2(\nu X)]^{\frac{1}{n-1}}$
  \vspace{0.1in} \\ [1ex]
 7  & $\ \ 1$ & $ \ \ \ \ \displaystyle\frac{2n\mu}{n+1}$  &  $\  \ \displaystyle \frac{n-1}{2n}$ & $n=m>1$ &  $[A\cosh^2(\nu X)]^{\frac{1}{n-1}}$
   \vspace{0.1in} \\ [1ex]
 8 & $\ \ 1$ & $\displaystyle-\frac{\mu (m+1)}{2}$ & $\displaystyle\frac{\sqrt{-\mu}\,(1-m)}{2}$ & $n=1,\, m<1$ & $[A\sinh^2(\nu X)]^{\frac{1}{1-m}}$
 \vspace{0.1in}\\ [1ex]
 9 & $\ \ 1$ & $\ \ \displaystyle \frac{\mu (m+1)}{2}$ & $\displaystyle\frac{\sqrt{-\mu}\,(1-m)}{2}$ & $n=1,\, m<1$ & $[A\cosh^2(\nu X)]^{\frac{1}{1-m}}$
\\ [2.0ex]
 \hline\hline
\end{tabular}
\end{table}

 Without loss of generality, we choose the condition $CX_x>0$ in Eqs.~(\ref{eq:7})-(\ref{eq:10}) as
 \bee \label{con}
  C=1, \qquad X_x>0.
  \ene
The self-similar  variable $X(x)$ admits an infinite choices such that the corresponding
exact solutions of GP$(m,n)$ equation (\ref{GP}) will display the
abundant structures. For the simple case $X(x)=ax+b$\ ($a>0$ since
$X_x>0$ is required), it follows from Eqs.~(\ref{eq:7})-({\ref{eq:10}) that all these
functions $\rho(x),\ g(x)$ and $v(x)$ reduce to the constants
and the obtained exact solutions (\ref{eq:9}) and (\ref{eq:10})
become the analytical travelling wave solutions which illustrate
envelope compacton solutions and spikon solutions of
GP$(m,n)$ equation (\ref{GP}) with constant coefficients~\cite{Yan6}. Here
we do not consider the travelling wave case, i.e,  $X(x)=ax+b$\ ($a>0$ since
$X_x>0$ is required). \\

\centerline{\bf\large 4. \ Wave propagations of envelope
solutions} \vspace{0.1in}

\centerline{\bf 4.1 \ Attractive nonlinearity $G<0$} \vspace{0.1in}

In this case we take $G=-1$. In the following we will choose some functions for the variable $X(x)$ to study the
wave propagations of the solutions (\ref{eq:9}) and (\ref{eq:10}) of Eq.~(\ref{GP}).

{\it Choice I of $X(x)$.} \ To consider the envelope compacton-like solutions of the GP$(2, 2)$ equation (\ref{GP}) is given by Eq.~(\ref{eq:9}) with $\Phi(X)$ defined by Case 1 in Table I. We here focus on $X=X(x)$ as the
Jacobi amplitude function
 \bee
  X(x)=2\ {\rm am}(x,k), \ \ \ {\rm for \ \ all} \ \ x\in\mathbb{R}
   \label{X1} \ene
which satisfies the required condition $X_x=2\,{\rm dn}(x,k)>0$, where ${\rm am}(x,k)$ is the  elliptic integrals of the Jaocibi
elliptic function ${\rm dn}(x,k)$ with the modulus $k\in (0, 1)$~\cite{DN}, i.e.
$${\rm am}(x,k)=\int^x_{x_0} {\rm dn}(s,k)ds, \qquad |{\rm am}(x,k)|\leq
\pi.$$
 Thus we have $|X(x)|\leq 2\pi$ in terms of Eq.~(\ref{X1}), which just leads to the
required region of the independent variable $|\nu X(x)|$ of
$\Phi(X)$, i.e.,
 \bee |\nu X(x)|=\left|\frac{1}{4}X(x)\right|=\left|\frac{1}{2}\,{\rm am}(x,k)\right|\leq \frac{\pi}{2},  \ \
  {\rm for \ \ all} \ \ x\in\mathbb{R}
 \label{eq:12} \ene
Moreover, for the $X(x)$ defined by Eq.~(\ref{X1}), the nonlinearity
$g(x)$ and external potential $v(x)$ of the GP$(n,n)$ equation are
rewritten as
  \bee \begin{array}{l}
  v(x)=-\mu [2\ {\rm dn}(x,k)]^{7/4}-\omega,  \vspace{0.1in} \cr
  g(x)=\displaystyle-\frac{1}{4}\Big[15\,{\rm dn}^2(x,k)+3(1-k^2)\,{\rm nd}^2(x,k)+k^2-2\Big],  \end{array}
    \label{Condition}\ene
where ${\rm nd}(x,k)=1/{\rm dn}(x,k)$.

We require that the compacton solution $\Phi(X)$ of the SGP$(n,n)$
equation (\ref{OGP}) is non-zero ($[A\cos^2(\nu
X)]^{\frac{1}{n-1}}\not=0$) only in its one period $(|\nu X|\leq
\pi/2)$ nearby the origin and zero for all other region for $X$,
but the intensity $|\psi_1(x,t)|^2$ related to the corresponding
solution (\ref{eq:9}) of GP$(n,n)$ is only a function of $x$. It
follow from Eqs.~(\ref{X1}) and (\ref{eq:12}) that the compacton-like
solution of the GP$(n,n)$ equation is nontrivial for all $x\in\mathbb{R}$,
which is very different from properties of the usual compacton
solutions~\cite{R1}, because of the choice of $X(x)$, which is related to the potential $v(x)$
 and nonlinearity $g(x)$ (see Eq.~(\ref{eq:7})).

 To illustrate the effect of $X(x)$ in the
compacton-like solution (\ref{eq:9}),  The compacton solution of
the SGP$(n,n)$ equation (\ref{OGP}) given in Case 1 of Table I is
displayed in Fig. 1(a) with respect to the variable $X$ not $x$.
The corresponding envelope compacton solutions of GP$(n,n)$ equation (\ref{GP}) is a
non-travelling wave solution for $x$
\bee\label{solu1}
\psi(x,t)=\frac{1}{\sqrt[2n]{2\,{\rm dn}(x,k)}}\left\{-\frac{2n\mu}{n+1}\cos^2\left[\frac{n-1}{n}\,{\rm am}(x,k)\right]^{1/(n-1)}\right\}e^{i\omega t},
\ene
 and the intensity
$|\psi(x,t)|^2$ is illustrated in Fig. 1(b) with respect to $x$. It
is easy to see that the top part of the profile has so much
changes when the modulus $k$ closes to $1$, which is different
from the usual compacton solution illustrated in Fig. 1(a), since
$X(x)$ is the Jacobi amplitude function of $x$ given by Eq.~(\ref{X1}).

\begin{figure}[h]
\begin{center}
\hspace{0.1in}
  {\scalebox{0.35}[0.3]{\includegraphics{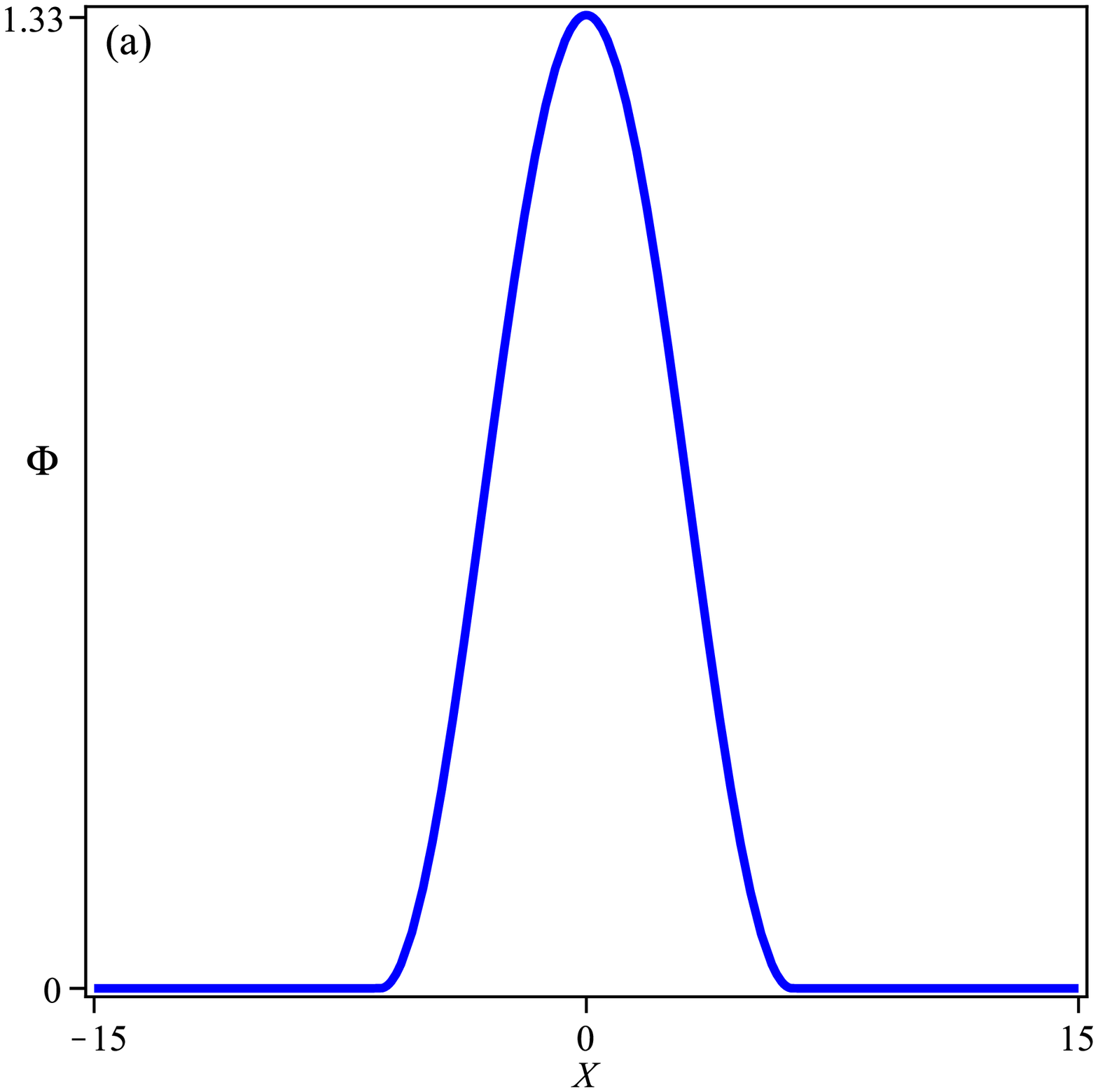}}}
  \hspace{0.1in}
  {\scalebox{0.35}[0.3]{\includegraphics{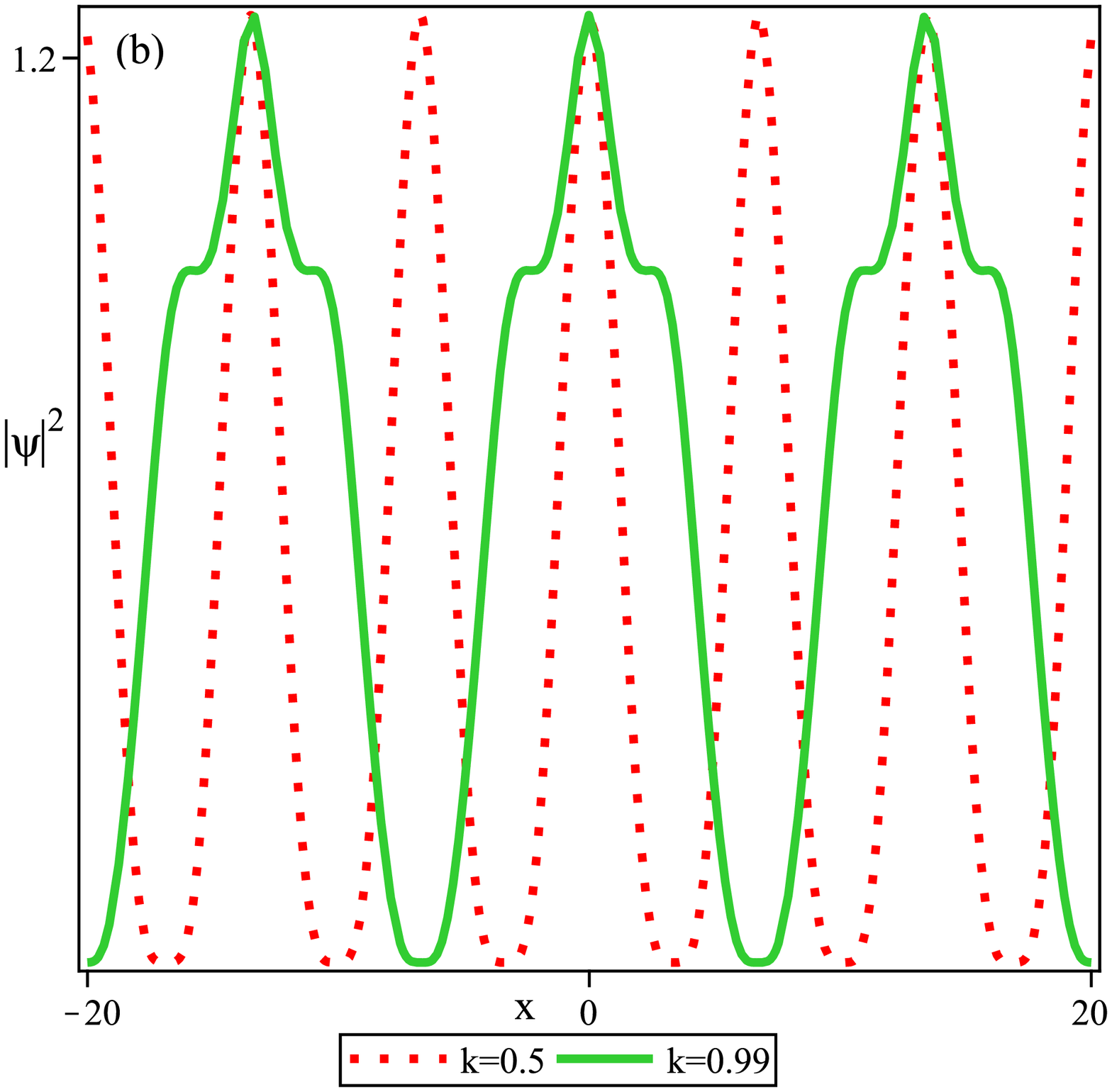}}}
  \end{center}
\vspace{-0.2in}\caption{\small Plots of $\Phi(X)$ and $|\psi|^2$ with the parameters are $n=2,\ \mu=-1$. (a)  $\Phi(X)$ given by Case 1 in Table I, (b) $|\psi(x,t)|^2$ given by Eq.~(\ref{solu1}) with $k=0.5$ and
$k=0.99$.}
\end{figure}

\begin{figure}[h]
\begin{center}
\hspace{0.1in}
  {\scalebox{0.35}[0.3]{\includegraphics{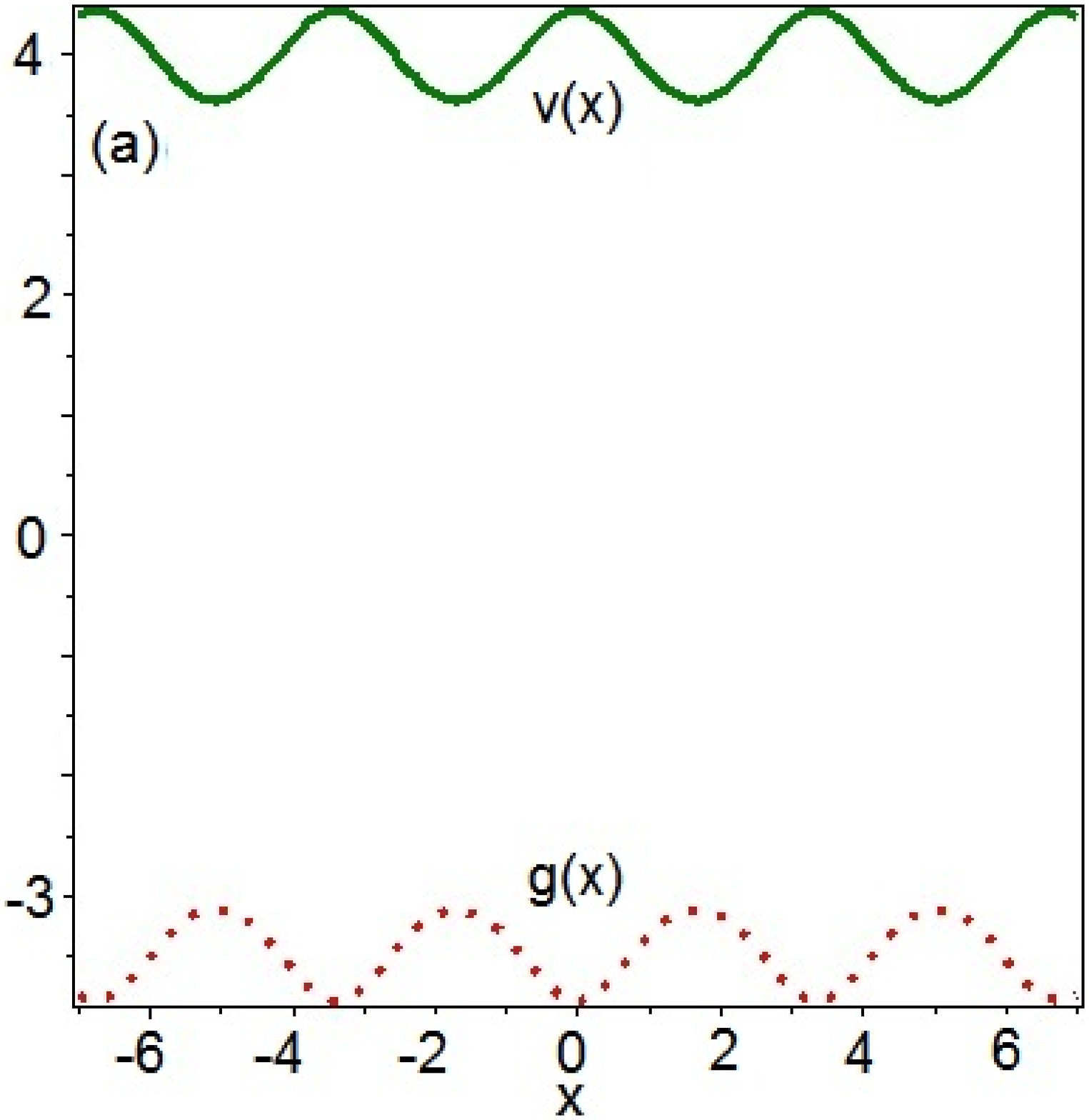}}} \hspace{0.1in}
   {\scalebox{0.35}[0.3]{\includegraphics{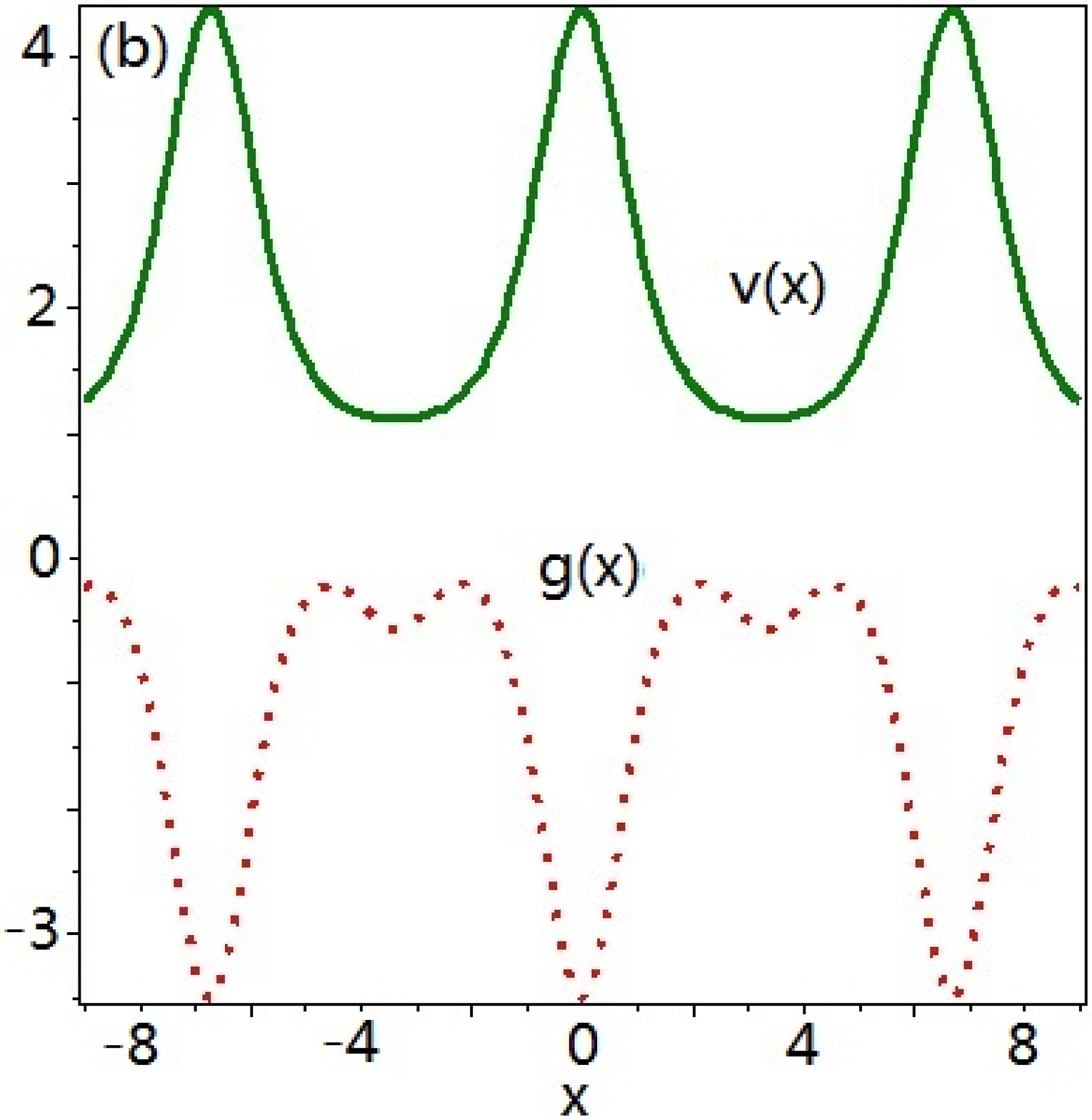}}}
\end{center}
\vspace{-0.2in}\caption{\small Plots of $g(x)$ and $v(x)$ given by Eq.~(\ref{Condition}) with $\omega=-1$ and othere parameters being the
same as Figure 1. (a) $k=0.5$,  (b) $k=0.99$.}
\end{figure}

In addition, the corresponding nonlinearity $g(x)$ and the
external potential $v(x)$ given by Eq.~(\ref{Condition}) are also
localized periodic waves and illustrated in Figs. 2(a,b). Of
course we can also choose other functions about $X(x)$ to illustrate
more types of solution structures.

For another case $n=1$ and $m<1$, in which GP$(m,1)$ becomes the
linear dispersive GP equation. Without loss of generality, we
choose $m=1/3,\ \mu=1$ and consider the exact solution
(\ref{eq:10}) of $GP(1/3, 1)$ equation with $\Phi(X)$ given by
Case 2 in Table I. We still choose the similarity variable $X(x)$
as the Jacobi amplitude function in the form
 \bee X(x)=\frac{3}{2}\ {\rm am}(x,k),
   \ \ \ {\rm for \ \ all} \ \ x\in\mathbb{R}
 \label{eq:13} \ene
which make sure that the following condition of the compacton
solutions of SGP$(1/3,1)$ equation
 \bee |\nu X(x)|=\left|\frac{1}{3}X(x)\right|=\left|\frac{1}{2}\,{\rm am}(x,k)\right|\leq \frac{\pi}{2},  \ \
  {\rm for \ \ all} \ \ x\in\mathbb{R}
 \label{Eq:14} \ene
holds. Moreover, for the $X(x)$ defined by Eq.~(\ref{X1}), the nonlinearity
$g(x)$ and external potential $v(x)$ of the GP$(n,n)$ equation are
rewritten as
  \bee \begin{array}{l}
  v(x)=\displaystyle\frac{1}{4}\Big[(1-9\mu)\,{\rm dn}^2(x,k)+3(k^2-1)\,{\rm nd}^2(x,k)-k^2+2\Big]-\omega,  \vspace{0.1in} \cr
  g(x)=\displaystyle -\left[\frac{3}{2}\,{\rm dn}(x,k)\right]^{5/3},  \end{array}
    \label{Condition2}\ene
where ${\rm nd}(x,k)=1/{\rm dn}(x,k)$.

In Fig. 3(a), we display the plots of the intensity
$|\psi(x,t)|^2$ of the solution given by Eq.~(\ref{eq:10}) of GP$(m, 1)$ equation in the form
\bee\label{solu2}
\psi(x,t)=\frac{\sqrt{2/3}}{\sqrt{{\rm dn}(x,k)}}\left\{-\frac{\mu(m+1)}{2}\cos^2\left[\frac{3\sqrt{\mu}(1-m)}{4}\,{\rm am}(x,k)\right]^{3/2}\right\}e^{i\omega t}, \,\, (m<1)
\ene
for different modulus
$k=0.5$ and $k=0.99$. When $k$ closes to $1$ from $0.98$, the top
part of the intensity profile appears the strong oscillation and
there exist more branches. Thus it follows from the conclusions
mentioned-above that the linear dispersive GP equation (GP$(m,1)$
equation) also admits the compacton-like solutions. Fig. 3(a) displays the profiles of the external potential and nonlinearity given by
Eq.~(\ref{Condition2}).

\begin{figure}[h]
\vspace{0.15in}
\begin{center}
\hspace{0.15in}
   {\scalebox{0.35}[0.3]{\includegraphics{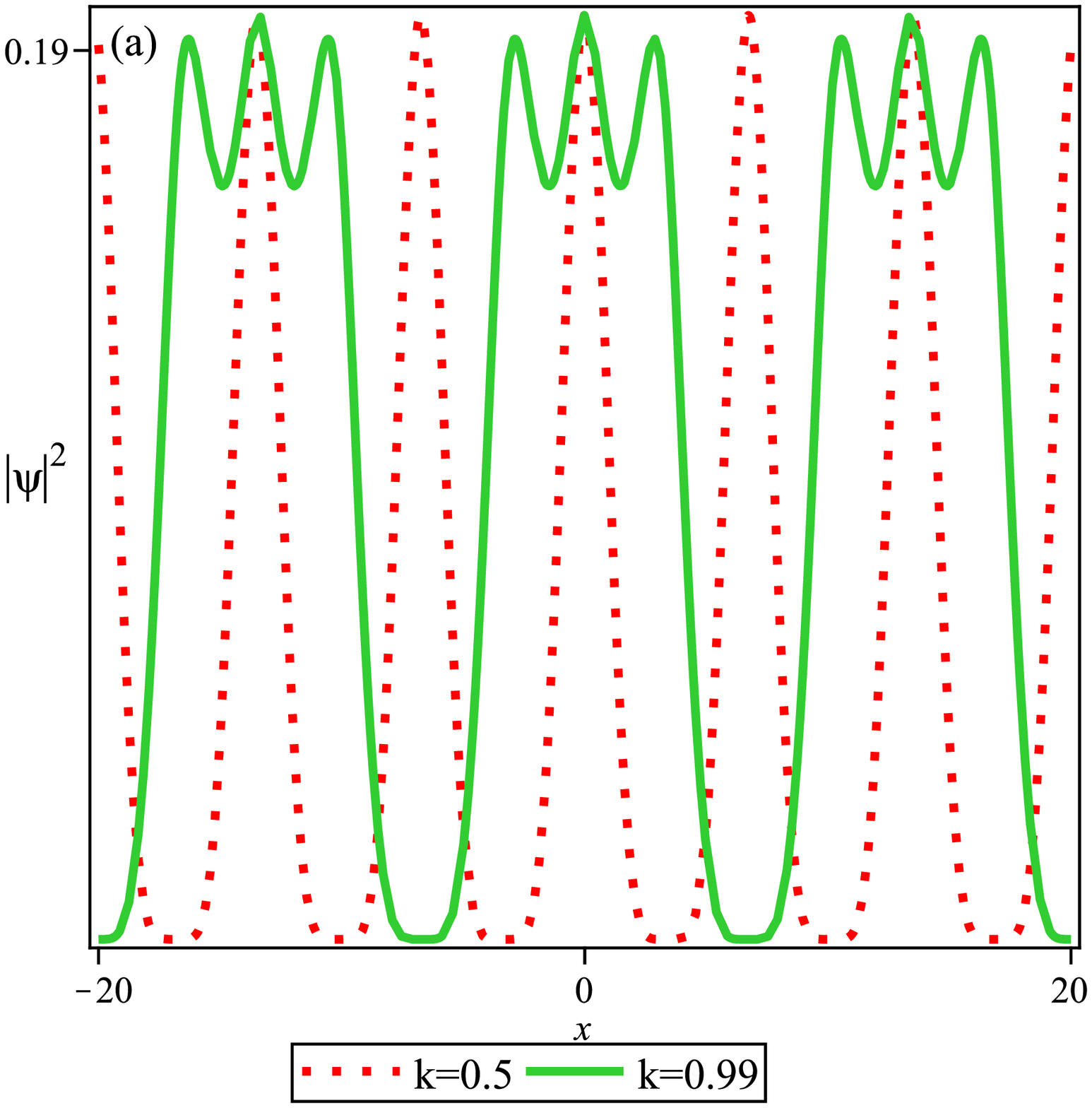}}} \hspace{0.1in}
   {\scalebox{0.35}[0.3]{\includegraphics{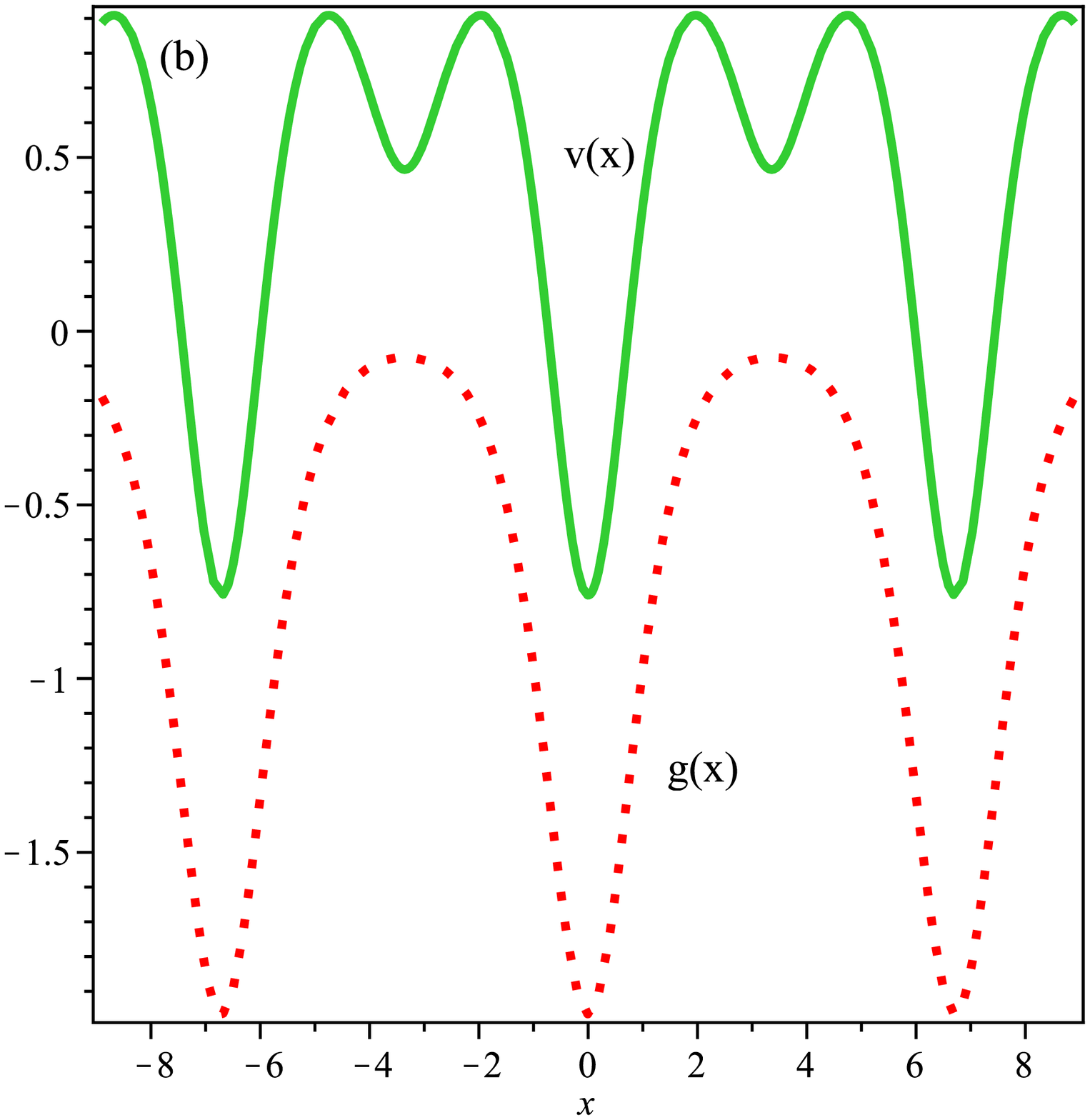}}}
\end{center}
\vspace{-0.2in}\caption{\small Solution (\ref{eq:10}) with $\Phi$
given in Case 2 of Table I with $X$ given by Eq.~(\ref{eq:13}) for $m=1/3,\ \mu=1$. (a) Plot of
$|\psi|^2$ given by Eq.~(\ref{solu2}) for $k=0.5,\ 0.99$, (b) Plots of $g(x)$ and $v(x)$
given by Eq.~(\ref{Condition2}) with $k=0.99$.}
\end{figure}

{\it Choice II of $X(x)$.} \ Here we consider another choice of
$X(x)$ in the form of the combination of linear and trigonometric
functions of $x$
 \bee X(x)=\alpha x+\sin(Kx), \label{C2} \ene
where $\alpha>K>0$ leads to the condition
$X_x=\alpha+K\cos(Kx)>0$. For the compacton-like
solutions of the GP$(2, 2)$ equation (\ref{GP}) is given by Eq.~(\ref{eq:9}) with $\Phi(X)$ defined by Case 1
in Table I, we require that $x$ satisfies the condition ($n=m=2$)
  \bee |\nu X(x)|=\left|\frac{1}{4}X(x)\right|=\left|\frac{1}{4}[\alpha x+\sin(Kx)]\right|\leq \frac{\pi}{2},
  \label{eq:17} \ene
For the given parameters $\alpha$ and $K$, one can find the region of $x$
which is a very small part of $(-\infty, \infty)$, distinguishing
from the condition (\ref{eq:12}). The corresponding envelope compacton solutions of GP$(n,n)$ equation (\ref{GP}) is a
non-travelling wave solution for $x$
\bee\label{solu3}
\psi(x,t)=\frac{1}{\sqrt[2n]{\alpha+K\cos(Kx)}}\left\{-\frac{2n\mu}{n+1}\cos^2\left[\frac{n-1}{2n}\,(\alpha x+\sin(Kx))\right]^{1/(n-1)}\right\}e^{i\omega t},
\ene

In Fig. 4, we find that more
strongly nonlinear wave oscillates, smaller the value
$\beta=\alpha/K-1$ becomes. When $\beta=0.05$, the wave profile is
divided into three parts being of three vertexes. Moveover, when
$\beta$ closes to $0$, the middle vertex increases slowly, but the
beside vertexes grow quickly. This is different from the usual
compacton solutions (see Fig. 1(a)).

\begin{figure}[h]
\begin{center}
\hspace{0.1in}
  {\scalebox{0.4}[0.3]{\includegraphics{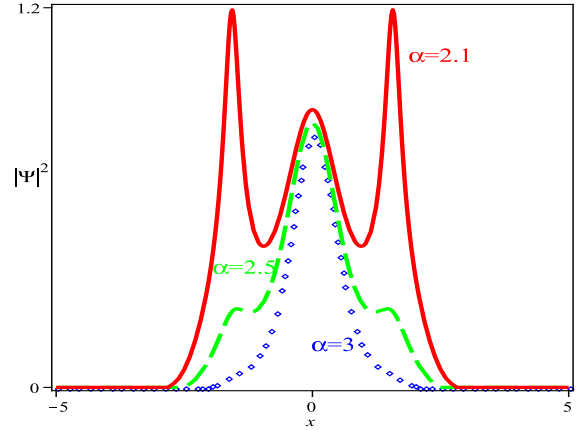}}}
  \end{center}
\vspace{-0.2in}\caption{\small Plots of $|\psi(x,t)|^2$ given by Eq.~(\ref{solu3})
 with the parameters $n=2,\ \mu=-1,\  K=2$ and $\alpha=3,\ 2.5,\ 2.1$.}
\end{figure}

In addition, based on Eqs.~(\ref{eq:7}) and (\ref{C2}), we have the corresponding nonlinearity and potential
in the form
 \bee
\begin{array}{l}
 v(x)=-\mu[\alpha+K\cos(Kx)]^{7/4}-\omega, \vspace{0.1in}\cr
 g(x)=\displaystyle \frac{3K^4\sin^2(Kx)}{4\left[\alpha+K\cos \left( Kx\right) \right]^2}
  +\frac{K^3\cos\left( Kx\right) }{ 2\left[\alpha+K\cos(Kx)\right]}
  -[\alpha+K\cos(Kx)]^{2},
  \end{array}  \label{eq:C2}
\ene which are illustrated in Figs. 5(a,b) for $\alpha=2.5,\, 2.1$,
and $K=2$.

\begin{figure}[h]
\vspace{0.15in}
\begin{center}
\hspace{0.1in}
   {\scalebox{0.35}[0.3]{\includegraphics{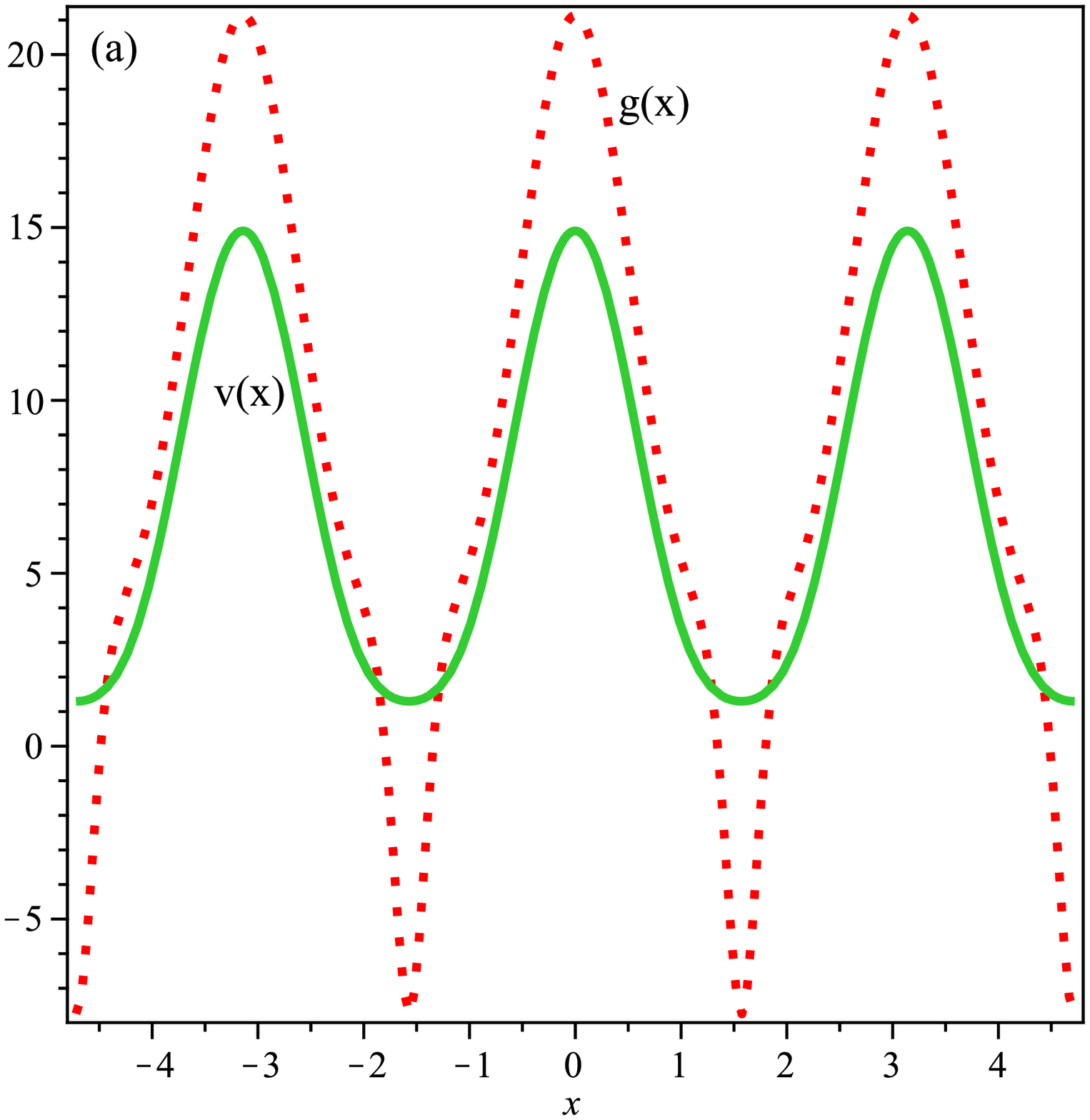}}} \hspace{0.1in}
   {\scalebox{0.35}[0.3]{\includegraphics{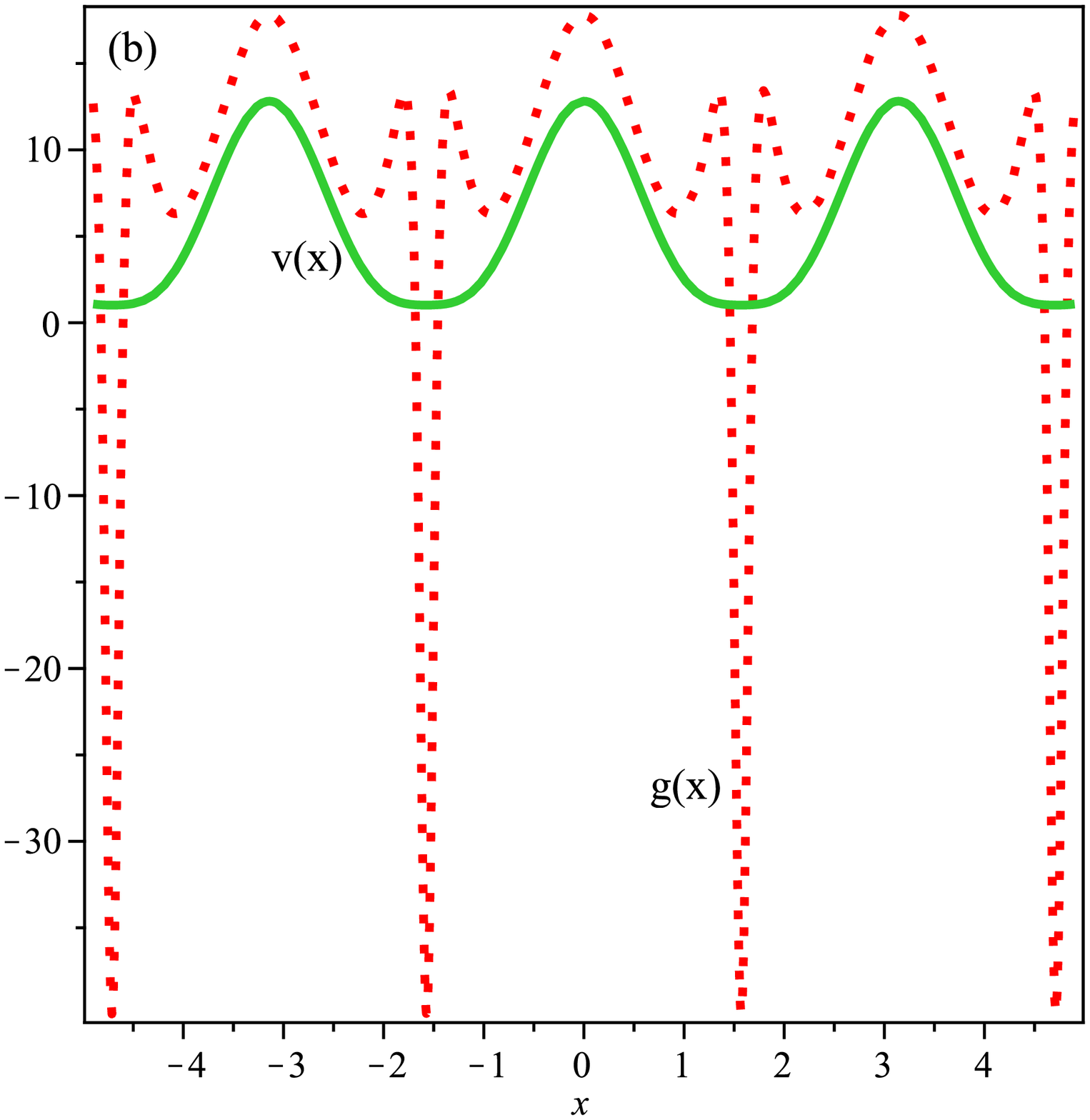}}}
\end{center}
\vspace{-0.2in}\caption{\small Plots of $g(x)$ and $v(x)$ given by
Eq.~(\ref{C2}) with $\omega=-1,\, \mu=-1, \, K=2$. (a) $\alpha=2.5$,
(b) $\alpha=2.1$.}
\end{figure}

For the case $n=1,\ 0<m<1$, where we consider the analytical exact
solution (\ref{eq:10}) of GP$(1/3, 1)$ equation with $\Phi(X)$
given by Case 2 in Table I. We still choose the similarity
variable $X(x)$ as (\ref{C2}) and we require that $x$ satisfies
the condition
  \bee |\nu X(x)|=\left|\frac{1}{3}X(x)\right|=\left|\frac{1}{3}[\alpha x+\sin(Kx)]\right|\leq \frac{\pi}{2},
  \label{eq:18} \ene
As a result, we have the envelope compacton-like solutions of GP$(m,1)$ equation in the form
\bee\label{solu4}
\psi(x,t)=\frac{1}{\sqrt{\alpha+K\cos(Kx)}}\left\{-\frac{\mu(m+1)}{2}\cos^2\left[\frac{\sqrt{\mu}(1-m)}{2}\,[\alpha x+\sin(Kx)]\right]^{3/2}\right\}e^{i\omega t}, \,\, (m<1)
\ene

In Fig. 6, we display the profile of the compacton-like solutions of GP$(m,1)$ equation and find that more strongly nonlinear wave oscillates,
smaller the value $\beta=\alpha/K-1$ becomes. When $\beta=0.04$,
the wave profile is divided into three parts being of three
vertexes. Moveover, when $\beta$ closes to $0$, the middle vertex
increases quickly, but the beside vertexes grow slowly. This is
also different from the usual compacton solution.

\begin{figure}[!h]
\begin{center}
\vspace{0.15in}
\hspace{0.3in}
  {\scalebox{0.4}[0.3]{\includegraphics{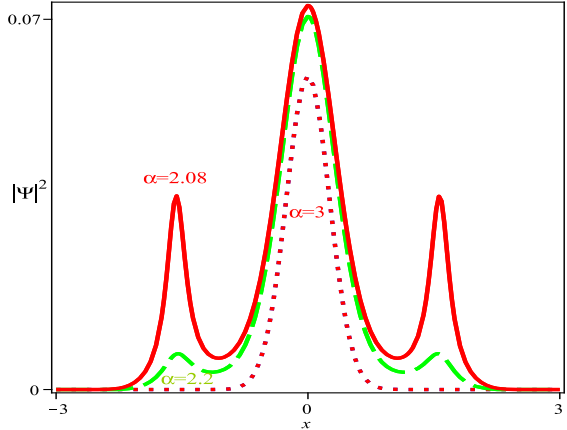}}}
  \end{center}
\vspace{-0.2in}\caption{\small Plots of $|\psi(x,t)|^2$ given by Eq.~(\ref{solu4}), corresponding to
$n=1,\ m=1/3,\ \mu=1,\ K=2$ and $\alpha=3,\ 2.2,\ 2.08$.}
\end{figure}

\vspace{0.1in}

 \centerline{\bf 4.2 \ Repulsive nonlinearity $G>0$} \vspace{0.1in}

 In this case we take $G=1$, the envelope spikon-like solutions of the GP$(2,2)$ equation
 (\ref{GP}) is given by (\ref{eq:9}) with $\Phi(X)$
defined by Case 7 of Table I. We still choose $X$ as the form
(\ref{X1}) even if we do not require the similar condition
(\ref{eq:12}). Though the hyperbolic cosh function $\cosh(X)$ is
infinite for $X$ approaches to infinity, we know that
\bee
 |X(x)|\leq
2\pi \ \ {\rm and} \ \ X_x=2\ {\rm dn}(x,k)\in (0,\, 2) \ \
  {\rm for \ \ all} \ \ x\in\mathbb{R}. \ene
 Thus the novel spikon-like solution
of GP$(n,n)$ equation is localized for all $x$ which is different
from the usual spikon solutions~\cite{R2,Yan, Yan6}.
Figure 7(a) denotes that the sipkon solution of SGP$(2,2)$
equation is not localized for the variable $X$, but the
spikon-like solution of GP$(2,2)$ is localized because of the
proper choice of $X(x)$ which is illustrated in Figure 7(b).
Similarly, if we choose $\Phi(X)$ as the Case 6 in Table I, then
the results are displayed in Figure 8.

For the case $n=1$ and $m<1$, without loss of generality, we
choose $m=1/3,\ \mu=-1$ and consider the solution (\ref{eq:10}) of
GP$(1/3, 1)$ equation with $\Phi(X)$ given by Case 9 in Table I.
We choose the new variable $X(x)$ as the Jacobi amplitude function
(\ref{eq:13}). The profiles of intensity $|\psi(x,t)|^2$ are
illustrated in Figures 9(a,b).

\begin{figure}[h]
\begin{center}
\hspace{0.1in}
   {\scalebox{0.35}[0.3]{\includegraphics{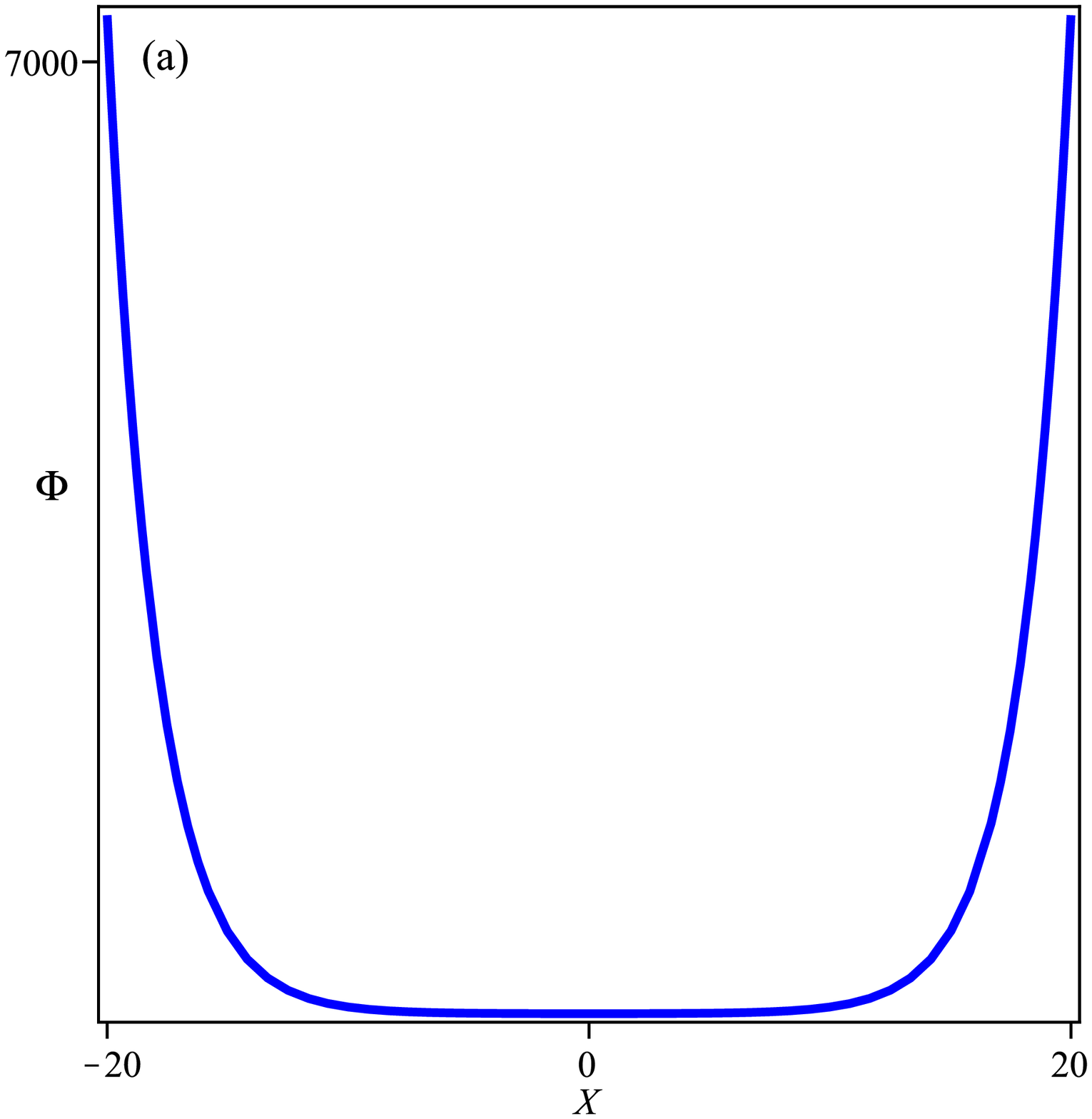}}} \hspace{0.1in}
   {\scalebox{0.35}[0.3]{\includegraphics{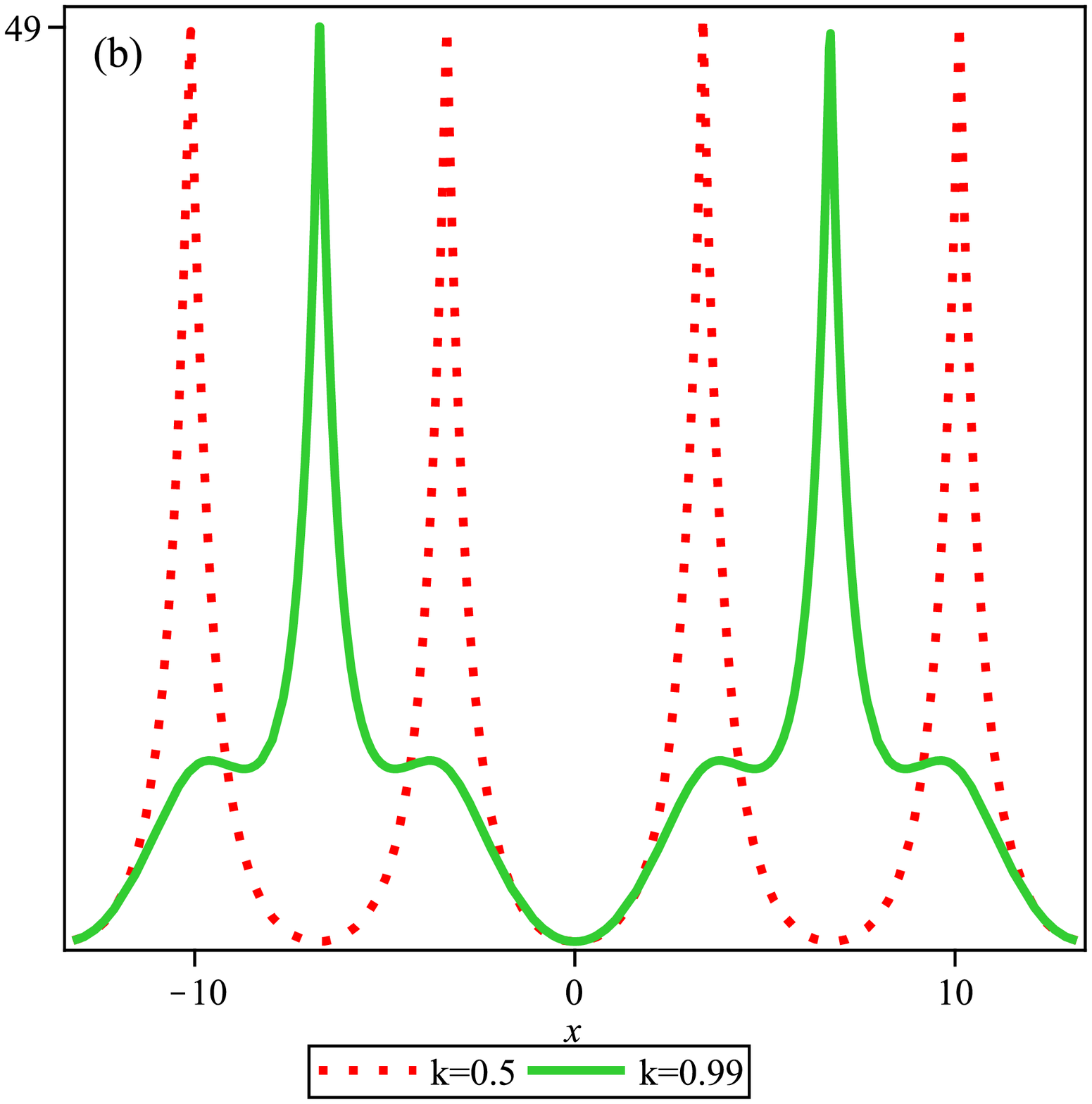}}}
\end{center}
\vspace{-0.2in}\caption{\small Solution given by Eq.~(\ref{eq:9}) with
$X(x)$ given by Eq.~(\ref{X1}) and $\Phi(X)$ given by Case 7 in Table
I, corresponding to (a) Plot of $\Phi(X)$ with $n=2,\ \mu=1$, (b)
Plots of $|\psi|^2$ with $k=0.5$ and $k=0.99$.}
\end{figure}

\begin{figure}[h]
\begin{center}
\hspace{0.1in}
   {\scalebox{0.35}[0.3]{\includegraphics{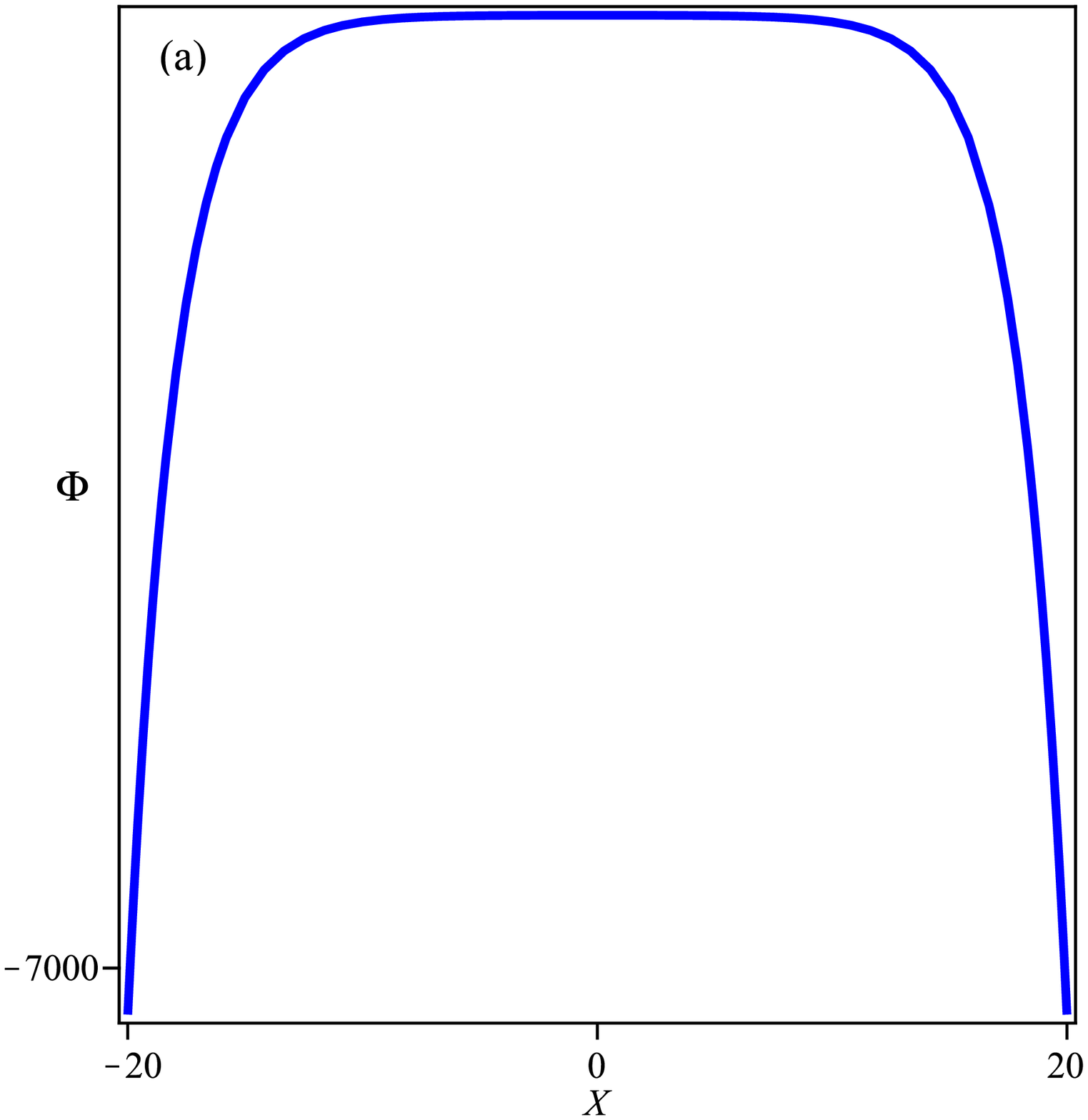}}} \hspace{0.1in}
   {\scalebox{0.35}[0.3]{\includegraphics{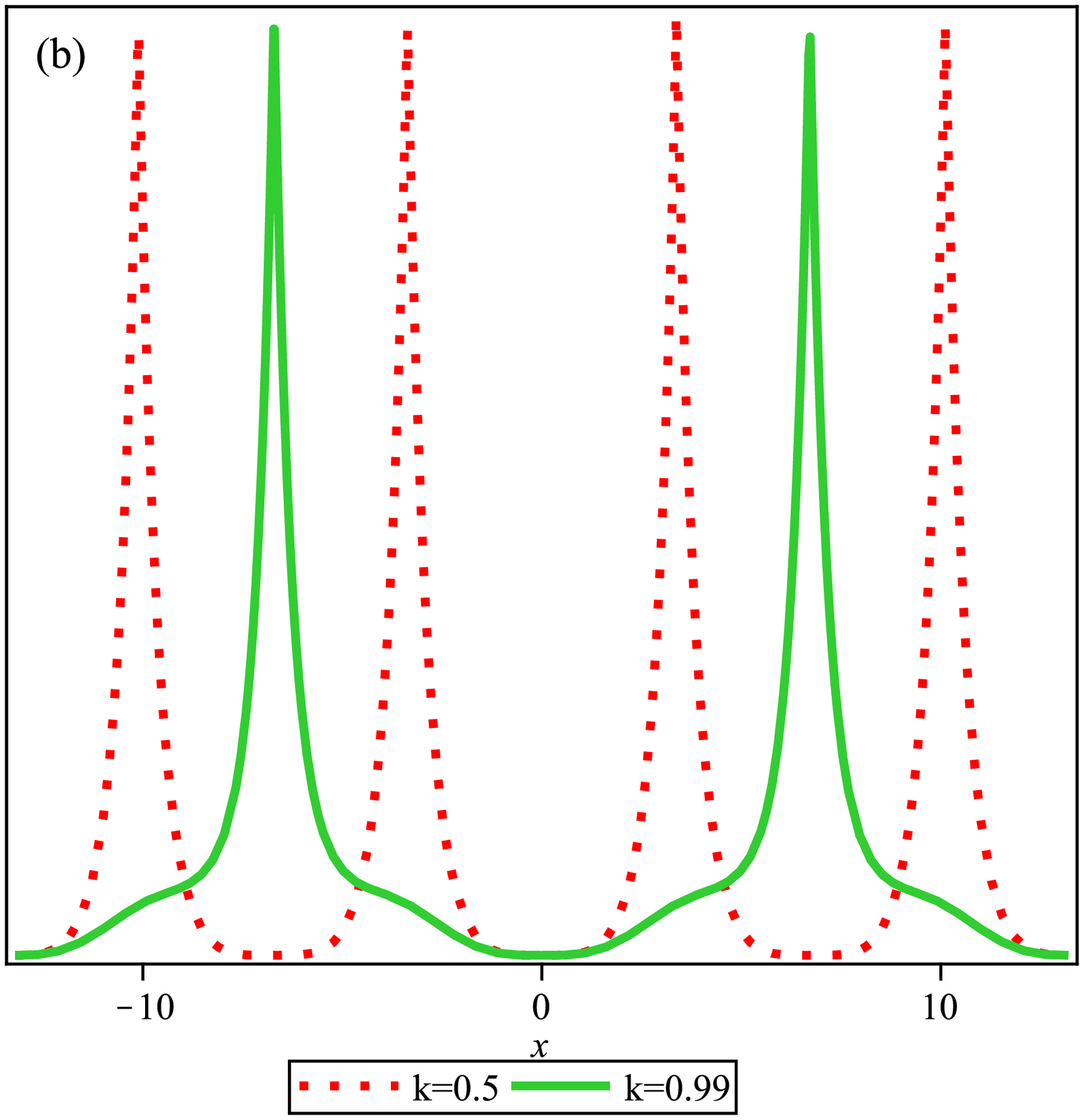}}}
\end{center}
\vspace{-0.2in}\caption{\small Solution given by Eq.~(\ref{eq:9}) with
$X(x)$ given by Eq.~(\ref{X1}) and $\Phi(X)$ given by Case 6 in Table
I, corresponding to (a) Plot of $\Phi(X)$ with $n=2,\ \mu=1$, (b)
Plots of $|\psi|^2$ with $k=0.5$ and $k=0.99$.}
\end{figure}

\begin{figure}[h!]
\begin{center}
\hspace{0.02in}
   {\scalebox{0.35}[0.3]{\includegraphics{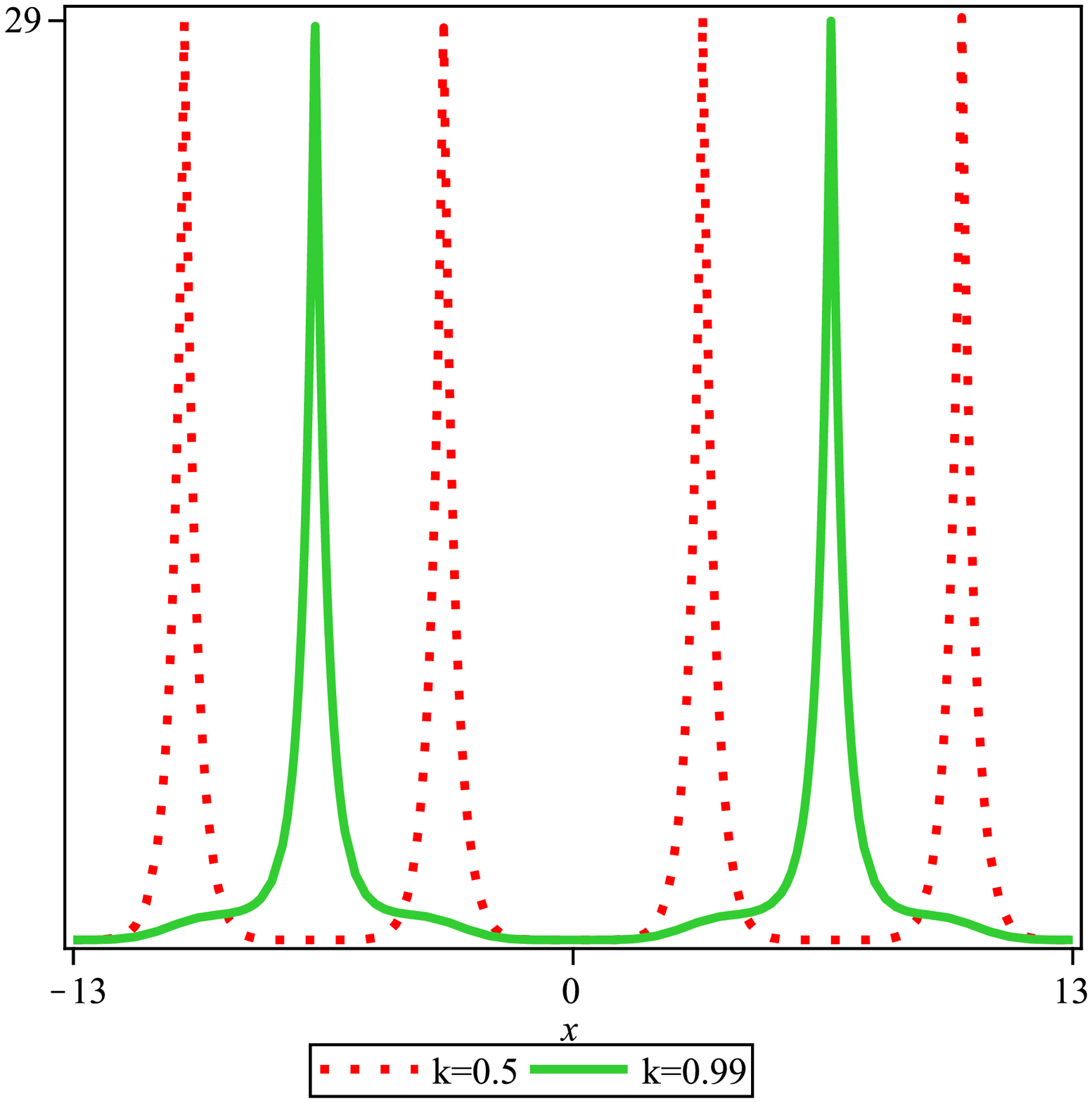}}} \hspace{0.1in}
   {\scalebox{0.35}[0.3]{\includegraphics{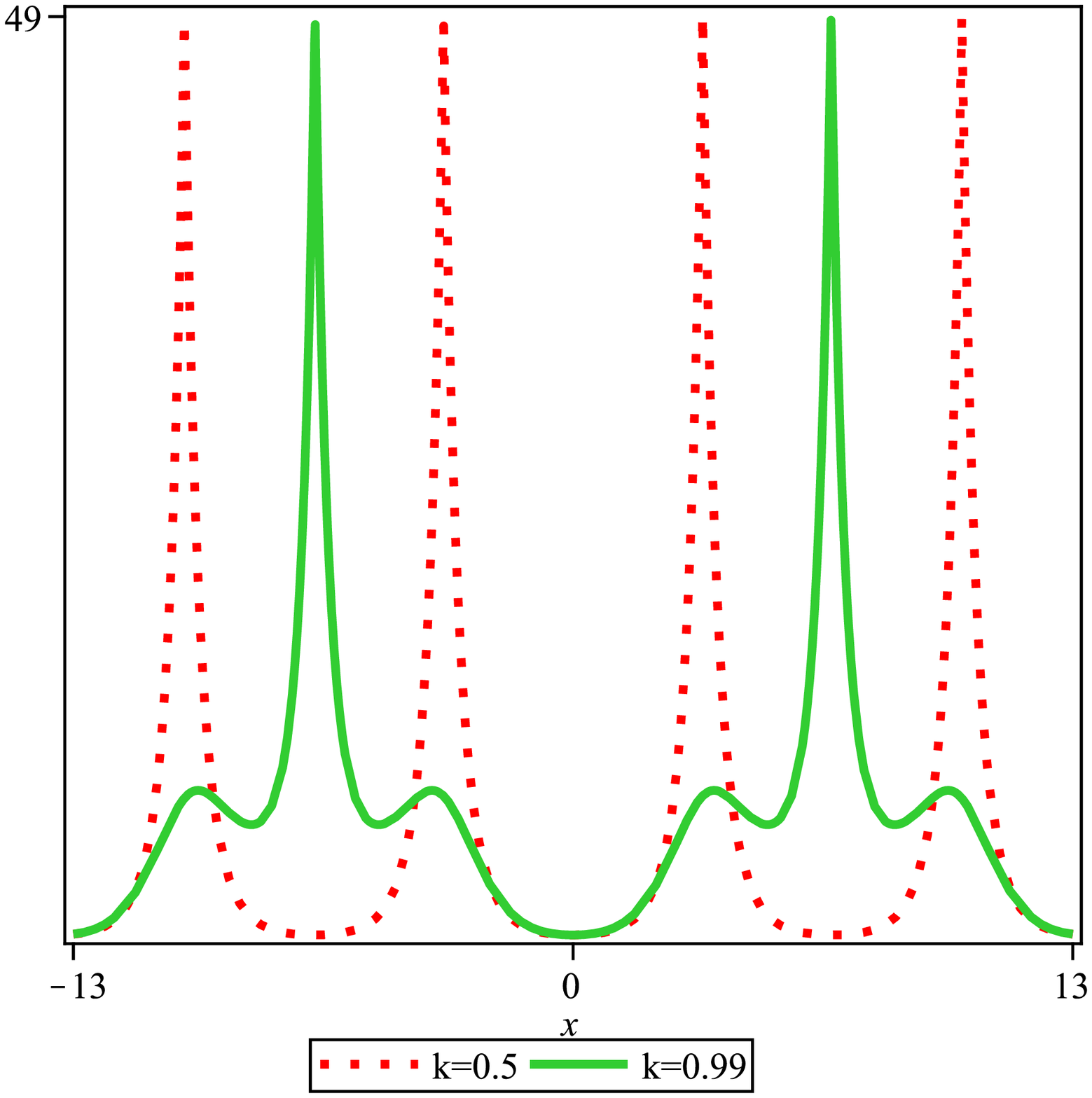}}}
\end{center}
\vspace{-0.2in}\caption{\small Solution given by (\ref{eq:9}) with
$X(x)$ given by (\ref{X1}) and $\Phi(X)$ given by Case 6 in Table
I, corresponding to (a) Plot of $\Phi(X)$ with $n=1,\, m=1/3,\, \mu=-1$, $k=0.5$ and $k=0.99$, (b)
Plots of $|\psi|^2$ with $k=0.5$ and $k=0.99$.}
 \end{figure}

In addition, it follows from Case 4 and 5 of Table I that the
GP$(m,1)$ equation with $G<0$ also admits the spikon-like
solutions which is similar to the case of GP$(n,n)$ equation with
$G>0$. On the contrary according to Case 3 of Table I, we know
that the GP$(m,1)$ equation with $G>0$ also admits the
compacton-like solutions which is similar to the case of GP$(n,n)$
equation with $G<0$.

Based on the above-mentioned results, we have the following proposition:

\noindent {\bf Proposition.} The nonlinear dispersive GP$(n,n)$ equation with $n>1$ possess the compacton-like solutions and spikon-like solutions. But nonlinear dispersion is not a necessary condition for the GP$(m,n)$ equation to possess these types of solutions. The linear dispersive GP$(m,1)$ equation also possess the compacton-like solutions and spikon-like solutions for $m<1$.

\centerline{\bf\large 5. \ Conclusions}

In summary, we have introduced the nonlinear dispersive GP$(m,n)$ equation with the space-modulated
nonlinearity and potential. By using some similarity
transformations and some powerful methods, we present some families of new exact
solutions with an arbitrary function $X(x)$ for different
parameters $m$ and $n$. It is shown that for the nonlinear
dispersion case $m=n>1$, we obtain some families of novel
compacton-like solutions and spikon-like solutions of
GP$(n,n)$ equation and that GP$(m,1)$ equation with linear
dispersion also admits similar solutions. That is to say,
nonlinear dispersion is not a necessary condition for nonlinear
wave equation to allow the compacton and spikon solutions.

 Although the function $X(x)$ admits the infinite kinds of choices, we focus on two special
cases of Jacobi amplitude function $X(x)=c\ {\rm am}(x,k)$ and the
combination of linear and trigonometric functions of $x$ to
investigate the obtained solutions. In particular, for the case of
Jacobi amplitude function, the corresponding compaton-like
solutions can nontrivially exist on the $\mathbb{R}$ and the
spikon-like solutions are shown to be localized. In addition, the
nonlinearity $g(x)$ and potential $v(x)$ are both the localized
period wave functions. These solutions may be useful to explain
some physical phenomena.

Moreover the idea can be also extended to other nonlinear
dispersive equations with varying coefficients such as the generalized GGP$(n,p,q)$ equation
 \begin{eqnarray}
 \nonumber i\frac{\partial \psi}{\partial t}
  =-\frac{\partial^2}{\partial x^2} (|\psi|^{n-1}\psi)+v(x)\psi+g_p(x)|\psi|^{p-1}\psi+g_q(x)|\psi|^{q-1}\psi,
  \label{GPg}
 \end{eqnarray}
and the three-dimensional GP$(m,n)$ equation
\begin{eqnarray}
 \nonumber i\frac{\partial \psi}{\partial t}=-\nabla^2(|\psi|^{n-1}\psi)+v(x,y,z)\psi
  +g(x,y,z)|\psi|^{m-1}\psi,
 \end{eqnarray}
where $\nabla^2=\partial_x^2+\partial_y^2+\partial_z^2$, which will be considered in the future.

\vspace{0.2in}\centerline{\large\bf  Acknowledgements}

  \vspace{0.1in}\noindent This work was partially supported by the
   NSFC of China (Nos. 11071242, 61178091).

\vspace{0.5in}\centerline{\sc Academy of Mathematics and Systems Science, Chinese Academy of Sciences}

 \end{document}